\documentclass[manuscript]{aastex62}

\newcommand{\del}{\partial}

\newcommand{\be}{  \begin{eqnarray} }
\newcommand{\ee}{  \end{eqnarray}} 
\newcommand{\bd}{  \begin{displaymath} }
\newcommand{\ed}{  \end{displaymath} }
\newcommand{\alphah}{{(a)}}
\newcommand{\betah}{{(b)}}
\newcommand{\gammah}{{(c)}}
\newcommand{\deltah}{{(d)}}
\newcommand{\ih}{{(i)}}
\newcommand{\rh}{{(r)}}
\newcommand{\thetah}{{(\theta)}}
\newcommand{\phih}{{(\phi)}}

\newcommand{\that}{{(t)}}

\newcommand{\zerohat}{{(0)}}
\newcommand{\onehat}{{(1)}}
\newcommand{\twohat}{{(2)}}
\newcommand{\threehat}{{(3)}}

\newcommand{\athenapp}{{\tt Athena++}}
\newcommand{\fnu}{n_\nu}
\newcommand{\fzeta}{n_\zeta}
\newcommand{\fpsi}{n_\psi}

\shortauthors{Davis \& Gammie}
\shorttitle{}

\begin{document}

\title{Covariant Radiative Transfer for Black Hole Spacetimes}

\author[0000-0001-7488-4468]{Shane W. Davis}
\affil{Department of Astronomy, University of Virginia}

\author{Charles F. Gammie}
\affil{Department of Astronomy, University of Illinois}

\begin{abstract}

It has now become possible to study directly, via numerical simulation, the evolution of relativistic, radiation-dominated flows around compact objects.  With this in mind we set out explicitly covariant forms of the radiative transfer equation that are suitable for numerical integration in curved spacetime or flat spacetime in curvilinear coordinates.  Our work builds on and summarizes in consistent form earlier work by Lindquist, Thorne, Morita and Kaneko, and others.  We give explicitly the basic equations in spherical-polar coordinates for Minkowski space and the Kerr spacetime in Kerr-Schild coordinates.

\end{abstract}

\section{Introduction}


Accreting black holes and neutron stars are among the most luminous objects in our galaxy.  Accreting supermassive black holes outshine their host galaxies and provide important probes of the high redshift universe.  The radiation and outflows they generate may play a role in the evolution of their parent galaxies \citep{2013ARA&A..51..511K}.  Accretion and associated outflows likely produce the electromagnetic counterparts to neutron star-neutron star merger events \citep{2017ApJ...848L..12A}.   Outflows from supermassive black holes are the likely accelerators for energetic particles that ultimately produce observed neutrinos from blazars \citep{1995APh.....3..295M}.   In both neutron stars and black holes most of the accretion energy is released in a relativistic region with Newtonian potential $\sim c^2$.  All this strongly motivates a numerical treatment of radiation in the relativistic regime and - because radiation stress can dominate other stresses in rapidly accreting objects - radiation hydrodynamics.  

Numerical relativistic radiation hydrodynamics (or magnetohydrodynamics) has received relatively little attention, however, because it is complicated and, relatedly, has high computational cost.  Advances in computational capabilities are on the cusp of making this interesting problem tractable.  This paper proposes a fully relativistic treatment of radiative transfer that is suitable for incorporation into radiation hydrodynamics schemes.

Let us begin by offering a few {\em desiderata} for a numerical scheme that will then determine the form in which we ultimately write the transfer equation.  First, we would like to write the basic equations in a form that is as close as possible to the familiar nonrelativistic transport equation, which in time dependent form can be written
\begin{equation}\label{eq.nonreltrans}
\left(\frac{1}{c}\frac{\partial}{\partial t} + {\bf n} \cdot \nabla\right) I_\nu = j_\nu - \kappa_\nu I_\nu.
\end{equation}
Here the fundamental dependent variable describing the radiation is the intensity $I_\nu = dE/dAdtd\nu d\Omega$ ($E \equiv$ energy, $A \equiv$ area, $\Omega \equiv$ solid angle), with cgs units erg cm$^{-2}$ sec$^{-1}$ ster$^{-1}$ Hz$^{-1}$.  Also, $j_\nu$ is the emissivity, and $\kappa_\nu$ is the extinction coefficient; these would be called collision terms in kinetic theory.  This equation is not invariant under Lorentz boosts or general coordinate transformations (covariant): $I_\nu, j_\nu$ and $\kappa_\nu$ are frame dependent, as is the derivative operator.   The derivative is a form of the ``Liouville operator'' or ``streaming term'' that forms part of the Boltzmann equation. This term is a consequence of two facts: (1) particle (here photon) number is conserved in the absence of collisions, and (2) photons motion through phase space is described by a Hamiltonian (again in the absence of collisions).  

Notice that we have already made an approximation in describing the radiation field using $I_\nu$ rather than the full set of Stokes parameters; see \citet{BroderickBlandford2004} \citet{Dexter2016}, \citet{MoscibrodzkaGammie2018}, \citet{GammieLeung2012} for recent covariant treatment of polarized transport theory.  For higher luminosity accretion flows, in which the dominant absorption absorption and emission processes are nearly independent of polarization, this should be a reasonable approximation.

There are many possible ways to rewrite the radiative transfer equation, but given common practice in the community we would like to continue to use $I_\nu$ to describe the radiation field.  This in turn requires specification of a set of preferred frames in which $I_\nu$ is defined.   

Second, we would like to write the transport equation in conservation form in both position and momentum space, so that it can be treated using common finite volume methods and so the resulting method conserves (e.g.) photon number.  Equation (\ref{eq.nonreltrans}) is written as an advection equation, not as a conservation law.   

Third, we would like to write the transport equation in as coordinate-independent way as possible.  That is, we do not wish to exploit any special symmetries of the system or special properties of a particular spacetime from the outset.

Fourth, we would like the scheme (and therefore the basic equations) to be as close as possible to those used in existing radiative transfer schemes.  This enables reuse of existing code, and a greater likelihood of having a correct and maintainable codebase.

So far, fully covariant integration of the radiation hydrodynamics equations in general relativity has taken one of two forms.  In the case of optically thin to moderately optically thick flows, Monte Carlo methods have been used with some success \citep[e.g.][]{Ryanetal2015}.  Monte Carlo methods generate unbiased samples of the radiation field (super photons or photon packets) that propagate through the simulation domain and interact with the background plasma, providing sources and sinks of momentum and energy that couple the radiation field to a finite volume integration of the general relativistic HD (or MHD) equations.   Monte Carlo methods either do not solve the transfer equation explicitly or partially solve it as an ODE along a ray.  Monte Carlo methods are promising but have typically not been used to integrate flows with high optical depths and large radiation forces because sampling noise limits the accuracy of the scheme for any reasonable number of samples and because (at least for naive implementations) the method tends to concentrate photon samples in regions of high optical depth.  Nevertheless, methods to extend time dependent Monte Carlo \citep{Foucart2018} and related sampling methods \citep{2019arXiv190709625R} to optically thick problems have been developed.  

Alternatively, one can start from the conservation of radiation stress-energy and directly integrate the components of the radiation stress-energy tensor using standard finite volume methods.  These are known as moment methods since the equations are integrated over frequency and solid angle.  Moment methods require a closure relation, with M1 closure \citep{Levermore1984,DubrocaFeugeas1999,Gonzalezetal2007} being the most popular for general relativistic methods  \citep{Sadowskietal2013,McKinneyetal2014,Fragiletal2014,Takahashietal2016}. Moment methods are particularly attractive because they are faster than methods that directly solve for the full angular structure of the radiation field.  Nevertheless, the closure relation is an approximation designed to reproduce the full solution in certain limits (e.g., for M1, the optically thick limit or the optically thin limit far from a point source), but it can fail outside these limits \citep[e.g.][]{2014ApJ...797..103C} because moment methods represent the radiation field with a reduced number of degrees of freedom (four for M1).  Therefore, development of a fully covariant angle and frequency dependent radiative transfer schemes are appealing to test the accuracy of moment methods and explore problems where closure relations may be insufficiently accurate.

These considerations motivate the development of a fully covariant formulation of the radiative transfer equation to be solved using finite volume methods, treating momentum space coordinates on the same footing as space-like coordinates.  The radiative transfer equation is a form of the Boltzmann equation, so we begin with the covariant Boltzmann equation and derive a flux conservative form for the radiative transfer equations that is appropriate for a finite volume numerical solution.  

The covariant Boltzmann equation was considered by \citet[based on Tauber's thesis work]{TauberWeinberg1961} and Israel (1963), and the first published treatment (to the best of our knowledge) of covariant radiative transport is \citet{Lindquist1966}.  The fullest development of covariant transport theory that we are aware of is aimed at understanding neutrino transport in the context of the core collapse supernova problem \citep[e.g.][]{Cardall2013b,Nagakuraetal2014}.  Thus, several of the results derived below have been previously described in other works.  Nevertheless, experience suggests it is useful to review the derivation and provide a self-contained description of the general covariant radiative transfer equation before deriving a specialized equation for the black hole spacetimes of interest to us.

The formalism outlined here is the basis for a fully finite volume implementation of radiative transfer under development for the \athenapp\, astrophysical radiation magnetohydrodynamics code.  We will use the formalism outlined below to implement radiative transfer in curvilinear coordinates in a flat spacetime as well as black hole spacetimes.  Details of the implementation and tests will be provided in future papers.

The plan of this work is as follows.  In section \ref{definitions} we summarize our notation and provide definitions of key quantities.  In section \ref{radiationtransfer} we review various forms of the covariant radiative transfer equation.  We verify that these equations obey the correct conservation relations when integrated over momentum space, consider the specific limits of a gray approximation and multifrequency moment methods, and outline a method for handling the matter-interaction terms. In section \ref{applications} we apply the results to specific spacetimes, first confirming that we recover the correct relations for spherical coordinates in flat space-time and then deriving specific forms for the Kerr spacetime of primary interest to accreting black hole. We summarize our conclusions and provide a guide to the main results in section \ref{summary}.

\section{Definitions and notation}
\label{definitions}

Before writing the transfer equation it is useful to fix our notation and definitions for key quantitites.  In addition to four spacetime coordinates, we will need to define three additional momentum coordinates.  These are most easily defined relative to an orthonormal tetrad basis.  Hence, we will need to distinguish quantities defined in the coordinate basis from those defined in the tetrad basis.  Tetrad basis components will be denoted $e^\alpha_\alphah$.  Indices with Greek letters (e.g. $\alpha$) are used to denote coordinate basis indices and Roman letters inside (\,) (e.g. $\alphah$) are used to denote indices in the tetrad basis. In some instances, it will be convenient to distinguish between one timelike and three general spacelike coordinates.  In these cases, we will denote the timelike coordinate by $t$ and the spacelike coordinate by $x^i$.  Indices $i$ and $\ih$ are used to denote the spacelike quantities in the coordinate and tetrad bases, respectively.  We use the Einstein summation convention (summation over repeated indices) throughout.

We define $x^\alpha$ and $k^\alpha$ to be the spacetime position and momentum four vectors of the photons in the coordinate frame.  Photons obey the geodesic equation
\be
\frac{d k^\gamma}{d \lambda}+\Gamma^\gamma_{\alpha\beta} k^\alpha k^\beta=0,
\ee
where $\lambda$ is an affine parameter.  Inserting $k^\gamma = k^\gammah e^\gamma_\gammah$ and performing some straight-forward algebra leads to
\be
\frac{d k^\gammah}{d\lambda} +  \omega^\gammah_{\alphah\betah} k^\alphah k^\betah = 0,\label{eq:connect}
\ee
where $\omega^\alphah_{\betah \gammah}$ are the Ricci connection coefficients
\be
\omega^\alphah_{\betah \gammah} \equiv e^\alphah_\alpha e^\gamma_\gammah e^\alpha_{\betah;\gamma},
\label{eq:ricci}
\ee
following \citet{Lindquist1966}.  Hence, the Ricci coefficients play the role of the connection in the tetrad basis. The Ricci coefficients have symmetry properties that are more clearly expressed when the first index is lowered to give
\be
\gamma_{\alphah\betah\gammah}= \eta_{\alphah\deltah} \omega^\deltah_{\betah\gammah}.
\ee
Here, $\eta=\,$diagonal(-1,1,1,1) is the Minkowski spacetime metric.  We refer to $\gamma_{\alphah\betah\gammah}$ as the Ricci matrices to distinguish them from the Ricci connection coefficients. The Ricci matrices are antisymmetric in their first two indices.

Throughout this work we use a spherical polar representation of momentum space.  The variables will always be the frequency $\nu$, polar angle $\zeta$, and azimuthal angle $\psi$ (angles $\theta$ and $\phi$ are used exclusively for spacelike coordinates in spherical-polar representations). In the tetrad basis, we can define $k^\alphah = \nu (1, n^\ih)$, where $\nu$ is the photon frequency and $n_\ih$ is a unit vector with $n_\ih n^\ih = 1$. Hence, $\nu = \sqrt{k^\ih k_\ih}$. A spherical-polar form for the momentum space is completed with the definition of angular variables $\zeta$ (polar) and $\psi$ (azimuthal) via
\be
n^{\onehat} & = & \sin \zeta \cos \psi\nonumber \\
n^{\twohat} & = & \sin \zeta \sin \psi \nonumber \\
n^{\threehat} & = & \cos \zeta.\label{eq:momentum}
\ee
Hence, we have seven independent variables: four space time coordinates $x^\alpha$ and three momentum space variables $\nu$, $\zeta$, and $\psi$.

 Following standard conventions, we use a subscript $\nu$ to indicate that quantities are a function of frequency $\nu$.  Hence, to avoid confusion, we never use $\nu$ as an index for coordinate frame tensors.  Similarly, we never use $\omega$ to denote frequency to avoid confusion with the Ricci connection coefficients.  We represent the radiation field by the specific intensity $I_\nu$ rather than photon distribution function, although it is easy to switch between the two.  The matter interaction terms are the emissivity  $j_\nu$ and total extinction coefficient  $\alpha_\nu$ ($\alpha_\nu = \rho \kappa_\nu$, with density $\rho$ and opacity $\kappa_\nu$).

\section{The Covariant radiative transfer Equation}
\label{radiationtransfer}

The solution of the radiation transfer equation is equivalent to the solution of the Boltzmann equation for photons.  There have been a number of treatments of derivations of the general covariant Boltzmann equation or radiation transfer equations in the literature \citep[e.g][]{Lindquist1966,Thorne1981,MoritaKaneko1986,Cardall2013b,Shibata2014}.  For completeness, we outline two derivations in appendices \ref{davis} and \ref{gammie}.  The derivation in appendix \ref{davis} roughly follows standard conventions, explicitly deriving the covariant transfer equation and manipulating it into a flux conservative form.  The derivation in appendix \ref{gammie} provides a somewhat simpler and more intuitive derivation, emphasizing that the covariant transfer equation follows from the phase space conservation principle underlying the Boltzmann equation.

These derivations show that the fully covariant radiation transfer equation can be written in the following flux conservative form
\be
& & (n^\alpha I_\nu/\nu)_{;\alpha} +
\frac{\partial}{\partial \nu}\left( \fnu \frac{I_\nu}{\nu} \right)
-\frac{1}{\sin \zeta}\frac{\partial}{\partial \zeta}\left(
  \fzeta \frac{I_\nu}{\nu}\right)+  \frac{\partial}{\partial \psi}\left(\fpsi \frac{I_\nu}{\nu} \right)
 =\frac{j_\nu - \alpha_\nu I_\nu}{\nu},
\label{eq:fluxcon}
\ee
where we define
\be
\fnu & = & -\nu n^\alphah n^\betah \omega^{\zerohat}_{\alphah \betah},\nonumber\\
\fzeta & = & -n^\alphah n^\betah \left(\omega^{\threehat}_{\alphah \betah} - \omega^{\zerohat}_{\alphah \betah}
n^{\threehat}\right),\nonumber\\
\fpsi & = &  n^\alphah n^\betah\frac{\left( n^{\twohat}\omega^{\onehat}_{\alphah \betah} - n^{\onehat}
    \omega^{\twohat}_{\alphah \betah}\right)}{\sin^2\zeta}\label{eq:fluxes}
\ee
Hence, specification of the radiative transfer equation in an arbitrary spacetime only requires choosing a tetrad basis, evaluating the connection coefficients, and evaluating the three terms in equation (\ref{eq:fluxes}).

The left hand side of equation (\ref{eq:fluxcon}) is in flux conservative form in that all terms are inside partial derivative with respect to the phase-space coordinates.  This follows from the fact that the Boltzmann equation represents conservation of phase space number density in the absence of matter-interaction source terms on the right hand side of the equation, as discussed in Appendix \ref{gammie}.

For many applications one prefers an evolution equation for $I_\nu$ rather than $n^t I_\nu/\nu$. We obtain such an equation by first multiplying through by $\nu$ and bringing $\nu$ inside the $\partial/\partial \nu$ term to obtain
\be
(n^\alpha I_\nu)_{;\alpha}  + 
\frac{\partial}{\partial \nu}\left(\fnu I_\nu \right)
-\frac{1}{\sin \zeta}\frac{\partial}{\partial \zeta}\left( \fzeta
  I_\nu\right)
 +   \frac{\partial}{\partial \psi}\left( \fpsi I_\nu \right)-\frac{\fnu I_\nu}{\nu}
 = j_\nu - \alpha_\nu I_\nu.
\label{eq:fluxconalt}
\ee
Next divide by $n^t$, bringing $n^t$ inside the partial derivatives.  Since our immediate interest is stationary spacetimes, we now specialize to time-independent tetrads and metrics.  We find
\be
\frac{1}{c}\frac{\partial I_\nu}{\partial t} & + & \frac{1}{\sqrt{-g}}\frac{\partial}{\partial x^i}
\left(\frac{\sqrt{-g} n^i}{c n^t} I_\nu \right) +
\frac{\partial}{\partial \nu}\left(\frac{\fnu}{c n^t} I_\nu \right)
-\frac{1}{\sin \zeta}\frac{\partial}{\partial \zeta}\left(\frac{\fzeta}{c n^t} I_\nu\right)
+ \frac{\partial}{\partial \psi}\left(\frac{\fpsi}{c n^t} I_\nu \right)\nonumber\\
& + & \left[\frac{n^i}{c (n^t)^2}\frac{\partial n^t}{\partial x^i}-\frac{\fnu}{c \nu n^t}-\frac{\fzeta}{c (n^t)^2}
\frac{1}{\sin \zeta} \frac{\partial n^t}{\partial \zeta}+\frac{\fpsi}{c (n^t)^2}
\frac{\partial n^t}{\partial \psi}\right] I_\nu=\frac{j_\nu - \alpha_\nu I_\nu}{c n^t}.\label{eq:dinudt}
\ee
where superscript $i$ denotes spacelike coordinates.  Equation (\ref{eq:dinudt}) simplifies if we choose our tetrad basis so that the space components $e^t_\ih=0$.  Then $n^t=e^t_\zerohat$ is independent of $\zeta$ and $\psi$.  In that case, the term in square brackets multiplying $I_\nu$ reduces to $n^i(n^t)^{-2}\partial n^t/\partial x^i-\fnu/(c\nu n^t)$.  Note that this term vanishes for flat spacetime and some other special cases but does not generally vanish in curved spacetimes.

Another form of the transfer equation can be obtained by multiplying equation (\ref{eq:fluxconalt}) by $n_\beta$.  As shown in appendix \ref{secondmoment}, the resulting expression simplifies to
\be
(n^\alpha n_\beta I_\nu)_{;\alpha} +
\frac{\partial}{\partial \nu}\left(\fnu n_\beta I_\nu \right)
-\frac{1}{\sin \zeta}\frac{\partial}{\partial \zeta}\left( \fzeta n_\beta I_\nu\right)
+\frac{\partial}{\partial \psi}\left(\fpsi n_\beta I_\nu \right)
=n_\beta \left(j_\nu - \alpha_\nu I_\nu\right),\label{eq:momentsimp}.
\ee
This can be rewritten as
\be
\frac{\partial (n^t n_\beta I_\nu)}{\partial t} & + &
\frac{1}{\sqrt{-g}}\frac{\partial}{\partial x^i}\left(\sqrt{-g} n^i n_\beta I_\nu \right)
+\frac{\partial}{\partial \nu}\left(\fnu n_\beta I_\nu \right)
-\frac{1}{\sin \zeta}\frac{\partial}{\partial \zeta}\left( \fzeta n_\beta I_\nu\right)
+\frac{\partial}{\partial \psi}\left(\fpsi n_\beta I_\nu \right)\nonumber \\
& + & n^\alpha n_\delta \Gamma^\delta_{\alpha\beta} I_\nu  =n_\beta \left(j_\nu - \alpha_\nu I_\nu\right)
.\label{eq:momentsimp2}
\ee
The left hand side of equation (\ref{eq:momentsimp2}) is almost in flux-conservative form except for the term $n^\alpha n_\delta \Gamma^\delta_{\alpha\beta} I_\nu$.  However, whenever the metric has an ignorable coordinate $x^\beta$ (so that $\partial g_{\alpha\delta}/\partial x^\beta=0$), $n^\alpha n_\delta \Gamma^\delta_{\alpha\beta}=0$ and the corresponding term vanishes. For example, when $t$ is an ingorable coordinate, we have
\be
\frac{\partial (n^t n_t I_\nu)}{\partial t} & + &
\frac{1}{\sqrt{-g}}\frac{\partial}{\partial x^i}\left(\sqrt{-g} n^i n_t I_\nu \right)
+\frac{\partial}{\partial \nu}\left(\fnu n_t I_\nu \right)
-\frac{1}{\sin \zeta}\frac{\partial}{\partial \zeta}\left( \fzeta n_t I_\nu\right)
+\frac{\partial}{\partial \psi}\left(\fpsi n_t I_\nu \right)\nonumber \\
& = & n_t \left(j_\nu - \alpha_\nu I_\nu\right)
.\label{eq:fluxconenergy}
\ee
Since $(n^t)^2 (I_\nu/c) d\nu d\Omega = h \nu f d^3\mathbf{p}$, where $f$ is the photon distribution function, this equation implies energy conservation for stationary spacetimes.  

Equations (\ref{eq:fluxcon}), (\ref{eq:dinudt}), and (\ref{eq:fluxconenergy}) are useful starting points for a number of implementations.  If one desires a gray (frequency integrated) treatment, one integrates  (\ref{eq:dinudt}) over frequency and the $\partial/\partial \nu$ term vanishes, while $I_\nu$ is simply replaced by $I=\int I_\nu d\nu$ in all other terms.  Alternatively, one could formulate a frequency dependent moment method by integrating over solid angle.  Or, one can attack the full six dimensional problem directly.

Our ultimate goal is a six-dimensional finite volume numerical method, but we defer detailed discussion of numerical implementation to future papers.  We simply note that equations (\ref{eq:fluxcon}) and (\ref{eq:fluxconenergy}) have the advantage of being written in flux-conservative form. In contrast, equation (\ref{eq:dinudt}) is not generally in flux-conservative form, although it is for the specific choice of flat spacetime with a time-independent tetrad basis.  Construction of a numerical scheme centered on equation (\ref{eq:fluxconenergy}) would seem particularly advantageous for conserving energy in the code.  It also yields a matter-interaction term which has the common form $\propto (j_\nu + \alpha_\nu I_\nu)$ without the extra factor of $1/\nu$ that appears in equation (\ref{eq:fluxcon}).  This could be beneficial for gray opacity treatments and applications where electron scattering opacity dominates since $\alpha_\nu$ is nearly independent of $\nu$ in the Thomson limit.  A potential downside of equation (\ref{eq:fluxconenergy}) is the extra dependence on $n_t$, which is generally a function of both momentum and real space coordinates, but this added complexity may not present difficulties in a numerical method since the corresponding variables may only need to be computed once.

\subsection{Verification of Conservation Relations}

Radiation must obey several conservation relations, including conservation of stress energy and (in the absence of absorption and emission) conservation of photon number.  The relations arise when the radiative transfer (or Boltzmann equation) is integrated over momentum space.  One can use differential forms to elegantly show that the relativistic Boltzmann equation produces the correct conservation laws and we refer the reader to \citet{Cardall2003} for a geometric interpretation of the cancellation of terms found below.

Conservation of photon number (in absence of absorption and emission) requires that the current
\be
N^\alpha = \int\int n^\alpha \nu^{-1} I_\nu \, d\nu d\Omega
\ee
obey $(N^\alpha)_{;\alpha}=0$.  In the absence of absorption and emission, equation (\ref{eq:fluxcon}) is equivalent to
\be
& & (n^\alpha I_\nu/\nu)_{;\alpha} +
\frac{\partial}{\partial \nu}\left( \fnu \frac{I_\nu}{\nu} \right)
-\frac{1}{\sin \zeta}\frac{\partial}{\partial \zeta}\left(
  \fzeta \frac{I_\nu}{\nu}\right)+  \frac{\partial}{\partial \psi}\left(\fpsi \frac{I_\nu}{\nu} \right)=0.
\ee
Integrating this equation over frequency and solid angle eliminates terms which are total derivative with respect to momentum coordinates, leaving
\be
(N^\alpha)_{;\alpha}=0.
\ee

The stress energy tensor of the radiation fields is given by
\be
R^{\alpha \beta}=\int\int n^\alpha n^\beta I_\nu \, d\nu d\Omega.
\ee
Integration of equation (\ref{eq:momentsimp}) over frequency and solid angle yields $R^\alpha_{\beta;\alpha} = -G_\beta$. Here,
\be
G_\beta =\int \int n_\beta \left(\alpha_\nu I_\nu-j_\nu\right) \, d\nu d\Omega.
\ee
Raising the second index yields the standard expression for conservation of stress energy
\be
R^{\alpha\beta}_{\;;\alpha} = -G^\beta,\label{eq:stressenergy}
\ee
where $G^\beta$ is the radiation four-force density obtained by integrating the source terms \citep{Mihalas1984}.

\subsection{Gray Approximation}

In a gray approximation, one integrates $I=\int I_\nu d\nu$ instead of $I_\nu$. The $\nu$ dependence of equation (\ref{eq:dinudt}) is rather simple since neither the Ricci connection coefficients nor the $n^\alphah$ depend explicitly on $\nu$. Therefore, we obtain a gray approximation by integrating over $\nu$, replacing $I_\nu$ with $I$
\be
\frac{1}{c}\frac{\partial I}{\partial t} + \frac{1}{\sqrt{-g}}\frac{\partial}{\partial x^i}
\left(\frac{\sqrt{-g} n^i}{c n^t} I\right)
-\frac{1}{\sin \zeta}\frac{\partial}{\partial \zeta}\left(\frac{\fzeta}{c n^t} I\right)
+ \frac{\partial}{\partial \psi}\left(\frac{\fpsi}{c n^t} I \right)
+ \mathcal{S} I =\frac{1}{c n^t} \int (j_\nu - \alpha_\nu I_\nu)\, d\nu,\label{eq:gray}
\ee
where
\be
\mathcal{S} \equiv \frac{n^i}{c (n^t)^2}\frac{\partial n^t}{\partial x^i}-\frac{\fnu}{c\nu n^t}-\frac{\fzeta}{c (n^t)^2}
\frac{1}{\sin \zeta} \frac{\partial n^t}{\partial \zeta}+\frac{\fpsi}{c (n^t)^2}
\frac{\partial n^t}{\partial \psi}
\ee
is independent of frequency. This equation is exact.

That a gray prescription is even viable may seem counterintuitive given that one expects the frequency shifting present in curved spacetime to impact the radiative transfer.  Redshifting does enter through the presence of the term $I\mathcal{S}$ and coordinate dependent $n^t$, neither of which are present in a conventional treatment of flat spacetime. The main difficulty with a gray treatment is the matter-interaction source term, which is defined in the tetrad frame in equation (\ref{eq:gray}).

We describe a method for handling the source terms with the gray approximation in section \ref{sourceterms}.  The main limitation is that we must use mean opacities.  However, since we are not directly computing the frequency dependence of $I_\nu$ these frequency averaged opacities are only approximate.  Since opacities can vary significantly with frequency, different mean opacities (e.g. Rosseland, Planck) can differ by almost an order of magnitude.  For relativistic calculations, the frequency shifting that occurs as we boost from the tetrad frame to the comoving frame is also lost.  In cases where the opacities is strongly frequency dependent, this could lead to substantial errors.

\subsection{Moment Methods}

A common approximation in the numerical solution of radiative transfer is to integrate the angle integrated radiation moments directly.  In such schemes, one solves the stress energy tensor evolution using equation (\ref{eq:stressenergy}). As with all moment methods, this requires a closure method such as M1. As discussed above, these closure schemes may be less accurate in limits they were not explicitly designed to model but they are still appealing because they greatly reduce the number of degrees of freedom and therefore the computational cost relative to direct solution of the transfer along a large number of angles.

One can imagine problems where the frequency variation of the radiation field may be more important than its angular variation. This can happen when key opacities depend sensitively on frequency.  In this case one might want to use a multifrequency moment method, including multigroup methods that average the radiation quantity over multiple frequency ranges (groups).  One can motivate such a scheme from equation (\ref{eq:momentsimp}) by raising $n_\beta$ and integrating over solid angle (but not frequency).  Using
\be
\fnu = -\nu n^\alphah n^\gammah \omega^\zerohat_{\alphah\gammah}=\nu n^\gamma n^\alpha e^\zerohat_{\gamma;\alpha},
\ee
we find
\be
R^{\alpha \beta}_{\nu;\alpha} + e^\zerohat_{\gamma;\alpha}\frac{\partial}{\partial \nu}\left(\nu Q^{\gamma\beta\alpha}_\nu\right)
= -G^\beta_\nu, \label{eq:multigroup}
\ee
where
\be
Q^{\gamma\beta\alpha}_\nu \equiv \int n^\gamma n^\beta n^\alpha \, I_\nu \, d\Omega,
\ee
and
\be
G^\beta_\nu = \int n^\beta \left(\alpha_\nu I_\nu-j_\nu\right) \, d\Omega
\ee
We remind the reader that the subscript $\nu$ in these equations is not an index, but simply denotes frequency dependent quantities. This expression agrees with \cite{Cardall2013a}.  A multigroup method can be derived from this equation by averaging the frequency dependent quantities presented here over a finite number of frequency groups and using the $\partial/\partial \nu$ term to evaluate fluxes between groups due to redshifting.  Hence, a rigorous implementation of the multigroup method for the moment equation requires estimating a higher order angular moment to compute the flux in frequency space from one group to the next.

\subsection{Matter-interaction Source Terms}
\label{sourceterms}

We now outline a method for handling the matter-interaction terms, deferring a detailed discussion of numerical implementation to a future paper. The matter-interaction source term on the right hand side of equation (\ref{eq:fluxcon}) is most easily evaluated in the comoving frame, which is the primary strength of the comoving frame approach.  In principle, one can construct a comoving frame tetrad and solve the transfer equation in this frame.  However, resulting expressions now depend directly on the fluid four-velocity and are unwieldy, even in flat spacetime.  Hence, we do not present them here but instead refer the reader to previous work \citep{MoritaKaneko1986,1986CoPhR...3..127M,Cardall2005}.

Fortunately, one can still take advantage of the simplicity of the comoving frame source term even when the transport operator is evaluated in the tetrad frame. We first focus on the case without scattering.  We evaluate the frequency ratio
\be
\frac{\tilde{\nu}}{\nu} = \frac{ k^\mu u_\mu}{k^\mu e^{\that}_\mu} = \frac{n^\alphah u_\alphah}{n^\alphah e^t_\alphah},
\ee
where $u^\alpha$ is the fluid four-velocity, and then use the standard transformation properties to write
\be
j(\mathbf{n},\nu)-\alpha(\mathbf{n},\nu) I(\mathbf{n},\nu)=
\left(\frac{\nu}{\tilde{\nu}}\right)^2\left[
\tilde{j}(\mathbf{\tilde{n}},\tilde{\nu})-\tilde{\alpha}(\mathbf{\tilde{n}},\tilde{\nu})
\tilde{I}(\mathbf{\tilde{n}},\tilde{\nu})\right],
\ee
where we write $I_\nu$ as $I(\mathbf{n},\nu)$ to make the frame dependence more explicit. This expression can be used to evaluate the source term in equation (\ref{eq:dinudt}) for inclusion (e.g.) via operator splitting.

For a gray transfer treatment, we can integrate over frequency noting that
\be
\int d\nu \rightarrow \frac{\nu}{\tilde{\nu}}\int d\tilde{\nu}
\ee
to obtain
\be
\int j_\nu(\mathbf{n},\nu)-\alpha(\mathbf{n},\nu) I(\mathbf{n},\nu) d\nu &= &
\left(\frac{\nu}{\tilde{\nu}}\right)^3 \int
\tilde{j}(\tilde{\nu})-\tilde{\alpha}(\tilde{\nu}) \tilde{I}(\mathbf{\tilde{n}},
\tilde{\nu}) d\tilde{\nu}\nonumber \\
& = & \left(\frac{\nu}{\tilde{\nu}}\right)^3 \left( \tilde{j} -\tilde{\alpha} \tilde{I}(\mathbf{\tilde{n}})\right).
\ee

Finally, we note that
\bd
\tilde{I}(\mathbf{\tilde{n}}) =\left(\frac{\tilde{\nu}}{\nu}\right)^4 I(\mathbf{n})
\ed
and obtain
\be
\int j_\nu(\mathbf{n},\nu)-\alpha(\mathbf{n},\nu) I(\mathbf{n},\nu) d\nu=
\left(\frac{\nu}{\tilde{\nu}}\right)^3 \tilde{j} - \left(\frac{\tilde{\nu}}{\nu}\right)
\tilde{\alpha} I(\mathbf{n}).
\ee
With scattering opacity a more sophisticated treatment is required, but the same basic procedure can be followed: Lorentz transform to the comoving frame using $u^\alphah$, compute the source terms in the comoving frame, and transform back to the tetrad frame.  In the more general case, this will require interpolation of intensities at different angles and frequencies. Since any conceivable scheme will couple the radiative transfer to the fluid equations, one may want to treat these source terms in a locally implicit fashion even when transport is treated explicitly \citep[see e.g.][]{Jiang2014}.  A covariant implict update is more complicated than what we present here, but still potentially feasible for non-relativistic or moderately relativistic limits.  In the relativistic limit where beaming effects become large, multiple angular grids might be necessary to handle the large anisotropies that may arise \citep{Nagakuraetal2014}.

\section{Applications to Specific metrics and Coordinate Systems}
\label{applications}

We now proceed to evaluate equation (\ref{eq:dinudt}) for specific spacetimes. We first consider the illustrative example of spherical-polar coordinates in Minkowski spacetime and then evaluate the Kerr spacetime transfer equation in Kerr-Schild coordinates.  Generalizing to other spacetimes, all that is necessary is to compute the Ricci connection coefficients for a specific choice of tetrad and spacetime.  A useful compendium of spacetimes with Ricci matrices worked out for simple choices of tetrads is provided by \citet{MuellerGrave2009}.  

\subsection{Spherical-Polar Coordinates}

We first consider Minkowski spacetime in spherical-polar coordinates as it provides some intuition for the more complicated black hole spacetimes.  Our spacetime coordinates are $\mathbf{x} = (t,r,\theta,\phi)$ with line element $ds^2 = -c^2 dt^2 + dr^2+r^2 d\theta^2 + r^2 \sin^2 \theta d\phi^2$.

These coordinates suggest an orthonormal basis with non-vanishing components
\be
e^t_\that = \frac{1}{c}, \;
e^r_\rh = 1, \;
e^\theta_\thetah = \frac{1}{r}, \;
e^\phi_\phih = \frac{1}{r \sin \theta}.
\ee
The non-vanishing Christoffel symbols are
\be
\Gamma^r_{\theta\theta} = -r, \;
\Gamma^r_{\phi\phi}=-r\sin^2\theta, \;
\Gamma^\theta_{r\theta}=\frac{1}{r}, \;
\Gamma^\theta_{\phi \phi}=-\sin\theta \cos \theta, \;
\Gamma^\phi_{r \phi}=\frac{1}{r}, \;
\Gamma^\phi_{\theta \phi} = \cot \theta,
\ee
with symmetry in the lower two indices.  From these we can compute the
Ricci connection coefficients using equation (\ref{eq:ricci}). The non-vanishing matrices are
\be
\gamma_{\rh \thetah \thetah} = -\frac{1}{r}, \;
\gamma_{\rh \phih \phih} = -\frac{1}{r}, \;
\gamma_{\thetah\phih \phih}=-\frac{\cot \theta}{r},
\ee
with antisymmetry in the first two indices. We define the polar axis relative to $\mathbf{e}_\rh$ so that $\that=\zerohat$, $\rh = \threehat$, 
$\thetah=\onehat$ and $\phih=\twohat$. We then have
\be
n^\alphah n^\betah \omega^{\rh}_{\alphah \betah} & = & -\frac{\sin^2 \zeta}{r}\\
n^\alphah n^\betah \omega^{\thetah}_{\alphah \betah} & = & \frac{1}{r}\left(
\cos \zeta \sin \zeta \cos \psi - \cot \theta \sin^2 \zeta \sin^2 \psi \right)\\
n^\alphah n^\betah \omega^{\phih}_{\alphah \betah} & = & \frac{1}{r}\left(
\cot \theta \sin^2 \zeta \sin \psi \cos \psi + \cos \zeta \sin \zeta \sin \psi \right)
\ee
Inserting these relations into equation (\ref{eq:dinudt}) and noting that $n^t=1/c$ yields
\be
 \frac{1}{c}\frac{\partial I_\nu}{\partial t}+
\frac{1}{r^2} \frac{\partial(r^2 \cos \zeta I_\nu)}{\partial r} + \frac{1}{r \sin \theta}
\frac{\partial(\sin \theta \, \sin \zeta \cos \psi I_\nu)}{\partial \theta}+
\frac{1}{r \sin \theta}\frac{\partial (\sin \zeta \sin \psi I_\nu)}{\partial \phi} \nonumber \\
-\frac{1}{r \sin \zeta} \frac{\partial(\sin^2 \zeta \, I_\nu)}{\partial \zeta}
-\frac{\sin \zeta \cos \theta}{r \sin \theta} \frac{\partial (
\sin \psi \, I_\nu)}{\partial \psi} =j_\nu - \alpha_\nu I_\nu,\label{eq:spherical}
\ee
in agreement with previous derivations \citep[e.g.][]{Mihalas1984}. The partial derivatives with respect to angles $\zeta$ and $\psi$  in equation (\ref{eq:spherical}) represent fluxes of intensity from one angle to another. Since photons travel on straight rays in flat spacetime, these fluxes really account for the changes in the definition of the angles with respect to spatially varying coordinates axes. These fluxes are associated with tendency for any outgoing ray to become more parallel to the radial unit vector and ingoing rays to become more orthogonal to the radial unit vector.

\subsection{Kerr-Schild}

Our primary goal is an expression for radiative transfer in a black hole spacetimes.  We are generally interested in studying spinning black holes so we derive the equations for Kerr spacetime, with Schwarzschild obtained in taking the $a \rightarrow 0$ limit.  In principle we could work in Boyer-Lindquist coordinates, but these are singular on the horizon. Obtaining a set of equations that are regular on the horizon has proved advantageous in GRMHD simulations \citep[e.g.][]{Gammie2003} by allowing the inner boundary to be located inside the event horizon.  This also means that the $a \rightarrow 0$ limit of the equations below correspond to a version of Eddington-Finkelstein coordinates.

In Kerr-Schild, the coordinates are $\mathbf{x}=(t,r,\theta,\phi)$, where we use units so that $GM=c=1$. The line element is then
\be
ds^2 & = & -\left(1-\frac{2 r}{\rho^2}\right)dt^2 +\left(1+\frac{2 r}{\rho^2}\right)dr^2+\rho^2d\theta^2 +\sin^2 \theta\left[\rho^2+a^2 \sin^2\theta\left(1+\frac{2 r }{\rho^2}\right)\right]d \phi^2\nonumber \\
& & +\frac{4 r}{\rho^2} dr dt -\frac{4 a r \sin^2\theta}{\rho^2}
d\phi dt -2a\left(1+\frac{2 r}{\rho^2}\right)\sin^2\theta d\phi dr
\ee
where $\rho^2=r^2+a^2\cos^2\theta$.  For brevity, we do not report the non-vanishing Christoffel symbols and refer the reader to earlier work \citep[e.g.][]{Takahashi2008}. Our choice of tetrad is based on a locally non-rotating reference frame (LNRF)
\be
\mathbf{e}_\that & = &\left[
  \left(1+\frac{2 r}{\rho^2}\right)^{1/2}, -\frac{2 r}{\rho^2}\left(1+\frac{2 r}{\rho^2}\right)^{-1/2}, 0, 0\right],\nonumber\\
\mathbf{e}_\rh & = &\left[0, \frac{\sqrt{A}}{\rho^2}\left(1+\frac{2 r}{\rho^2}\right)^{-1/2}, 0,
  \frac{a}{\sqrt{A}}\left(1+\frac{2 r}{\rho^2}\right)^{1/2}\right],\nonumber\\
\mathbf{e}_\thetah & = &\left[0,0, \frac{1}{\rho},0\right],\nonumber\\
\mathbf{e}_\phih & = & \left[0, 0, 0,
  \frac{\rho}{\sqrt{A} \sin \theta}\right],
\ee
where $A=(r^2+a^2)^2-a^2\Delta \sin^2\theta$ and $\Delta = r^2-2 r+a^2$.  The relation to the LNRF is determined by our choice of $\mathbf{e}_\that$.  In principle, one could make other choices for $\mathbf{e}_\ih$, but the above vectors are chosen with the expectation that they will simplify expressions for the Ricci matrices.  Note that we use a slightly different set of coordinate and tetrad than those presented in \citet{Shibata2014}, who otherwise follow a similar approach for studying covariant transfer in the Kerr spacetime.

With these definitions, the non-vanishing Ricci matrices are
\be
\gamma_{\that \rh \that} & = & -\frac{\left(r^2-a^2 \cos^2 \theta\right)\sqrt{A}}{\rho^6\left(1+\frac{2 r}
  {\rho^2} \right)^{3/2}}, \nonumber\\
\gamma_{\that \rh \rh} & = &-
\frac{2 \left[r\rho^4(r^3+3 r^2 +a^2(r-1) \cos^2 \theta)-(2r^3 +a^2 \rho^2 \cos^2 \theta)A \right]}{\rho^6\left(1+\frac{2 r}{\rho^2}\right)^{3/2} A}, \nonumber\\
\gamma_{\that \rh \thetah} & = & \frac{2 a^2 r \cos \theta \sin \theta}{\rho^3
  \left(1+\frac{2 r}{\rho^2}\right) \sqrt{A}},\nonumber\\
\gamma_{\that \rh \phih} & = & -\frac{a \sin\theta\left[
    -3 r^4 + a^4 \cos^2\theta - a^2 r^2 (1 + \cos^2 \theta)\right]}{\rho^3 A},\nonumber\\
\gamma_{\that \thetah \that} & = &\frac{2 a^2 r  \sin \theta \cos \theta}{\rho^5 \left(1+\frac{2 r}
  {\rho^2}\right)},\nonumber\\
\gamma_{\that \thetah \thetah} & = &\frac{2 r^2 }{\rho^4 \left(1+\frac{2 r}{\rho^2}\right)^{1/2}},\nonumber\\
\gamma_{\that \thetah \phih} & = & -\frac{2 a^3 r  \sin^2 \theta \cos \theta}
 {\rho^4 \left(1+\frac{2 r}{\rho^2}\right)^{1/2}\sqrt{A}},\nonumber\\
\gamma_{\that \phih \phih} & = & \frac{2 r [r\rho^4-a^2\sin^2\theta (r^2-a^2\cos^2\theta)]}{\rho^4\left(1+\frac{2 r}{\rho^2}\right)^{1/2}A},\nonumber\\
\gamma_{\rh \thetah \rh} & = & -\frac{a^2\cos\theta\sin\theta\left[A+2 r(r^2+a^2)\left(1+\frac{2 r}{\rho^2}\right)\right]}{\rho^{3}\left(1+\frac{2 r}{\rho^2}\right) A},\nonumber\\
\gamma_{\rh \thetah \thetah} & = & -\frac{r \sqrt{A}}{\rho^4\left(1+\frac{2 r}{\rho^2}\right)^{1/2}},\nonumber\\
\gamma_{\rh \thetah \phih} & = & \frac{4 a^3 r^2 \cos\theta\sin^2\theta}{\rho^4
  \left(1+\frac{2 r}{\rho^2}\right)^{1/2}A},\nonumber\\
\gamma_{\rh \phih \phih} & = & -\frac{r\rho^4-a^2\sin^2\theta (r^2-a^2\cos^2\theta)}{\rho^4\left(1+\frac{2r}{\rho^2}\right)^{1/2} \sqrt{A}},\nonumber\\
\gamma_{\thetah \phih \phih} & = & - \frac{\rho^2A+2a^2r (r^2+a^2)\sin^2\theta}{\tan \theta \rho^3A},\label{eq:riccimatrices}
\ee
with $\gamma_{\that\thetah\rh} =\gamma_{\that \rh \thetah}$, $\gamma_{\that \phih \rh} =\gamma_{\that \rh \phih}$, 
$\gamma_{\that \phih \thetah} =\gamma_{\that \thetah \phih}$, $\gamma_{\rh \thetah \that} = \gamma_{\that \rh \thetah}$,
$\gamma_{\rh \phih \that} = - \gamma_{\that \rh \phih}$,
$\gamma_{\rh \phih \thetah}= \gamma_{\rh \thetah \phih}$, $\gamma_{\thetah \phih \that}= -\gamma_{\that \thetah \phih}$,
and $\gamma_{\thetah \phih \rh}= -\gamma_{\rh \thetah \phih}$.
The remaining non-zero elements can be found by noting antisymmetry in the first two indices.

Inserting (\ref{eq:riccimatrices}) into equations (\ref{eq:fluxcon}) or (\ref{eq:dinudt}) provides a complete description of the radiative transfer equation in the Kerr spacetime. The three momentum terms simplify to
\be
\fnu & = &\nu \cos \zeta\left[\gamma_{\that\rh\that}+\cos \zeta \gamma_{\that\rh\rh}
  +2 \sin \zeta \left(\cos \psi \gamma_{\that\rh\thetah}+\sin \psi \gamma_{\that\rh\phih}\right) \right]\nonumber\\
& & +\sin^2 \zeta \left[\cos^2\psi \gamma_{\that\thetah\thetah}+2\sin \psi\cos\psi\gamma_{\that\thetah\phih}+\sin^2\psi\gamma_{\that\phih\phih}
  \right]\nonumber\\
& &  +\sin \zeta \cos \psi \gamma_{\that\thetah\that}\nonumber\\
\fzeta  & = &
-\sin^2 \zeta \left[ \cos^2 \psi \gamma_{\rh\thetah\thetah}+2\sin\psi \cos \psi \gamma_{\rh\thetah\phih}+\sin^2\psi\gamma_{\rh\phih\phih} \right.\nonumber\\
  & & \left.-\left(\gamma_{\that\rh\that}+\cos \zeta \gamma_{\that\rh\rh}+2\sin\zeta\sin\psi\gamma_{\that\rh\phih}\right)\right .\nonumber\\
  & & \left. +\cos \zeta \left(\cos^2\psi \gamma_{\that\thetah\thetah}+2\sin \psi\cos\psi\gamma_{\that\thetah\phih}+\sin^2\psi\gamma_{\that\phih\phih}\right)\right]\nonumber\\
& & -\cos \zeta \sin \zeta \cos \psi \left(2 \cos \zeta  \gamma_{\that\rh\thetah} + \gamma_{\rh\thetah\rh}+\gamma_{\that\thetah\that}\right)\nonumber\\
\fpsi & = & \frac{\sin\psi}{\sin \zeta}
\left[-\gamma_{\that\thetah\that}-\gamma_{\rh\thetah\rh}\cos^2\zeta+\sin\zeta\cos\psi
  \left(\gamma_{\that\phih\phih}-\gamma_{\that\thetah\thetah}\right)
    \right .\nonumber\\
    & & \left. +\cos\zeta\sin\zeta\cos\psi \left(\gamma_{\rh\phih\phih}-\gamma_{\rh\thetah\thetah} \right)
    -2 \sin\zeta\sin\psi\gamma_{\that\thetah\phih}\right.\nonumber\\
    & & \left.-2 \cos \zeta \gamma_{\that\rh\thetah}+\sin^2 \zeta \gamma_{\thetah\phih\phih} -2\cos \zeta\sin\zeta\sin\psi
    \gamma_{\rh\thetah\phih}\right].
\ee
 Although the resulting expressions are unwieldy, they are a suitable starting point for a numerical method.  These quantities (or their integrals) will only need to be computed once, possibly numerically, at the beginning of a simulation run.
    

\section{Summary and Conclusions}
\label{summary}

We have built on the existing literature for solving the Boltzmann equation in curved spacetimes to derive fully covariant representations of the radiative transfer equation.  Although we are not the first to derive a general formulation, we provide simple expressions, intuitive derivations, and outline a description for how to apply this formalism to specific spacetimes and coordinate systems.   We first derive equation (\ref{eq:fluxcon}), which provides and evolution equation for the quantity $n^t I_\nu/\nu$, representing photon number conservation.  We also derive equation (\ref{eq:dinudt}), which is closely related but provides an evolution equation for $I_\nu$. It is only in flux-conservative form for flat spacetimes.  Finally, we provide an evolution equation (\ref{eq:fluxconenergy}) for $n^t n_t I_\nu$, which expresses energy conservation and can be written in flux-conservative form for spacetimes in which time is an ignorable coordinate.

After verifying conservation relations and considering frequency and angle averaged implementations, we evaluate the radiative transfer equation in some example spacetimes. We confirm that our formulation recovers previously derived results for spherical polar coordinates in flat spacetime and provide a formulation of the transfer equation in the Kerr spacetime.  These equations form the basis of a general relativistic, six dimensional finite volume scheme, which has been implemented in the \athenapp\ code and will described in a future paper.


\acknowledgements
We thank the annonymous referee for carefully reading the manuscript and providing helpful comments.  We are also grateful to Hiroki Nagakura for proving comments on our draft as well as Josh Dolence, Yan-Fei Jiang, Eliot Quataert, Ben Ryan, Jim Stone, Chris White, and the rest of the horizon collaboration for helpful discussions.
SWD acknowledges support from NASA Astrophysics Theory Program grant 80NSSC18K1018 and an Alfred P. Sloan Research Fellowship. CFG acknowledges support from NSF grant AST-1333612 and AST-1716327 and a Romano Professorial Scholarship.

\appendix

\section{Derivation of the Transfer Equation in a Tetrad Basis}
\label{davis}

Covariant treatments of the relativistic Boltzmann equation or radiation transfer equation have been provided by a number of authors \citep[e.g][]{Lindquist1966,Thorne1981,MoritaKaneko1986,Cardall2013b}. In simplest form, the covariant transfer equation is 
\be
\frac{D(I_\nu/\nu^3)}{d\lambda}= \frac{j_\nu -\alpha_\nu I_\nu}{\nu^2}.
\label{eq:transfer}
\ee
The quantity $I_\nu/\nu^3$ only differs from the photon distribution function by a constant factor and its transport is performed by the Liouville operator, which can be written as
\be
\frac{D}{d\lambda} = k^\alpha \frac{\partial}{\partial x^\alpha} + \frac{d k^\alpha}{d \lambda}\frac{\partial}{\partial k^\alpha},\label{eq:liouville}
\ee
where $x^\alpha$ and $k^\alpha$ are four vectors representing the spacetime position and momentum of the photons.  The partial derivatives $\partial/\partial k^\gamma$ in equation (\ref{eq:liouville}) must be evaluated on the light cone so there are only three independent momentum space coordinates.

Experience has shown that a spherical-polar representation of momentum space coordinates is effective for radiative transfer problems.  Following previous work, we define this representation relative to a tetrad basis who components can be written as $\mathbf{e}_\alphah$.  In the tetrad basis, the Liouville operator becomes
\be
\frac{D}{d\lambda} = k^\alphah e^\alpha_\alphah \frac{\partial}{\partial x^\alpha} + \frac{d k^\alphah}{d \lambda}\frac{\partial}{\partial k^\alphah}.
\ee
Substituting from equation (\ref{eq:connect}) we have
\be
\frac{D}{d\lambda} = k^\alphah e^\alpha_\alphah \frac{\partial}{\partial x^\alpha} -
\omega^\gammah_{\alphah \betah} k^\alphah k^\betah \frac{\partial}{\partial k^\gammah},\label{eq:lambda2}
\ee
which agrees with equation (95.44) of Mihalas \& Mihalas (1984). 

Using the spherical-polar representation in equation (\ref{eq:momentum}), we can replace the
derivatives with respect to $k^\alphah$ via
\be
\frac{\partial}{\partial k^\ih} = 
\frac{\partial \nu}{\partial k^\ih} \frac{\partial}{\partial \nu} +
\frac{\partial \zeta}{\partial k^\ih} \frac{\partial}{\partial \zeta} +
\frac{\partial \psi}{\partial k^\ih} \frac{\partial}{\partial \psi},
\ee
and
\begin{equation}
\begin{array}{lll}
\frac{\partial \nu}{\partial k^{\onehat}} = \sin \zeta \cos \psi, & 
\frac{\partial \zeta}{\partial k^{\onehat}} = \frac{\cos \zeta \cos \psi}{\nu}, &
\frac{\partial \psi}{\partial k^{\onehat}} = -\frac{1}{\nu} \frac{\sin \psi}{\sin \zeta},\\
\frac{\partial \nu}{\partial k^{\twohat}} = \sin \zeta \sin \psi, & 
\frac{\partial \zeta}{\partial k^{\twohat}} = \frac{\cos \zeta \sin \psi}{\nu}, &
\frac{\partial \psi}{\partial k^{\twohat}}= \frac{1}{\nu} \frac{\cos \psi}{\sin \zeta},\\
\frac{\partial \nu}{\partial k^{\threehat}} = \cos \zeta, & 
\frac{\partial \zeta}{\partial k^{\threehat}} = -\frac{\sin \zeta}{\nu}, &
\frac{\partial \psi}{\partial k^{\threehat}} = 0.\label{eq:partials}
\end{array}
\end{equation}
Collecting these terms we have
\be
\frac{d k^\alphah}{d \lambda}\frac{\partial}{\partial k^\alphah} =-
\frac{k^\alphah k^\betah}{\nu} & & \left [ n_\ih \omega^\ih_{\alphah \betah} 
\nu \frac{\partial}{\partial \nu} +\frac{\left(n_\ih \omega^\ih_{\alphah \betah} n^{\threehat} - 
  \omega^{\threehat}_{\alphah \betah} \right)}{\sin \zeta}\frac{\partial}{\partial \zeta}
\right. \nonumber\\
 & & \left. \;\; -\frac{\left( n^{\twohat}\omega^{\onehat}_{\alphah \betah} - n^{\onehat}
\omega^{\twohat}_{\alphah \betah}\right)}{\sin^2\zeta} \frac{\partial}{\partial \psi}
\right ].\label{eq:lambda3}
\ee 

We can rewrite $n_\ih \omega^\ih_{\alphah \betah}$ in terms of $\omega^{\zerohat}_{\alphah \betah}$ 
using the identity
\be
k^\alphah k^\betah n_\ih \omega^\ih_{\alphah \betah} =  k^\alphah k^\betah \omega^{\zerohat}_{\alphah \betah}.\label{eq:riccisum}
\ee
Inserting this into equation (\ref{eq:lambda3}), the Liouville operator takes the form
\be
\frac{D}{d\lambda} & = & k^\alphah e^\alpha_\alphah \frac{\partial}{\partial x^\alpha}
 -\frac{k^\alphah k^\betah}{\nu} \omega^{\zerohat}_{\alphah \betah} 
  \nu \frac{\partial}{\partial \nu}\nonumber\\
& &
+ \frac{k^\alphah k^\betah}{\nu} \left [ \frac{\left(\omega^{\threehat}_{\alphah \betah} -
\omega^{\zerohat}_{\alphah \betah} n^{\threehat}\right)}{\sin \zeta}\frac{\partial}{\partial \zeta}+
\frac{\left( n^{\twohat}\omega^{\onehat}_{\alphah \betah} - n^{\onehat}
\omega^{\twohat}_{\alphah \betah}\right)}{\sin^2\zeta} \frac{\partial}{\partial \psi}
\right ]\label{eq:lambda4},
\ee
which was previously derived by \citet{MoritaKaneko1986}).

At this point we have seven independent coordinates, the four spacetime coordinates $x^\alpha$, and three momentum space coordinates, for which we have adopted $\nu$, $\zeta$, and $\psi$.  Since these are independent variables, derivatives with respect to spacetime coordinates do not operate on $\nu$, $\zeta$, or $\psi$.  The three momentum space coordinates are defined relative to a tetrad basis and the spacetime variation of the tetrad basis has already been accounted for by the connection $\omega^\alphah_{\betah \gammah}$.

We now apply the operator in equation (\ref{eq:lambda4}) to $I_\nu/\nu^3$. Inserting into equation (\ref{eq:transfer}) and multiplying by $\nu$ we obtain
\be 
n^\alphah e^\alpha_\alphah \frac{\partial (I_\nu/\nu)}{\partial x^\alpha} -
n^\alphah n^\betah \omega^{\zerohat}_{\alphah \betah} \left(
\frac{\partial (I_\nu)}{\partial \nu} -\frac{3 I_\nu}{\nu}\right) 
+n^\alphah n^\betah \frac{\left(\omega^{\threehat}_{\alphah \betah} - 
\omega^{\zerohat}_{\alphah \betah} n^{\threehat}\right)}
{\sin \zeta}\frac{\partial (I_\nu/\nu)}{\partial \zeta}
\nonumber \\
+ n^\alphah n^\betah\frac{\left( n^{\twohat}\omega^{\onehat}_{\alphah \betah} - n^{\onehat}
\omega^{\twohat}_{\alphah \betah}\right)}{\sin^2\zeta} \frac{\partial (I_\nu/\nu)}{\partial \psi}
=\frac{j_\nu - \alpha_\nu I_\nu}{\nu}
\label{eq:general}
\ee

Since we are ultimately interested in a finite volume implementation, it is useful to rewrite equation (\ref{eq:general}) in a form that is closer to a flux conservative form.  The spacetime derivatives can be written in terms of a divergence
\be
n^\alphah e^\alpha_\alphah \frac{\partial (I_\nu/\nu)}{\partial x^\alpha}= (n^\alpha I_\nu/\nu)_{;\alpha}
-\frac{I_\nu}{\nu}
n^\alphah e^\alpha_{\alphah;\alpha}.
\ee
This, along with straightforward application of the chain rule yields
\be
& & (n^\alpha I_\nu/\nu)_{;\alpha} -
\frac{\partial}{\partial \nu}\left(n^\alphah n^\betah \omega^{\zerohat}_{\alphah \betah} I_\nu \right)
+\frac{1}{\sin \zeta}\frac{\partial}{\partial \zeta}\left[ n^\alphah n^\betah
  \left(\omega^{\threehat}_{\alphah \betah} - \omega^{\zerohat}_{\alphah \betah} n^{\threehat}\right)
  \frac{I_\nu}{\nu}\right]\nonumber\\
& & +  \frac{\partial}{\partial \psi}\left[ n^\alphah n^\betah\frac{
    \left( n^{\twohat}\omega^{\onehat}_{\alphah \betah} - n^{\onehat}
    \omega^{\twohat}_{\alphah \betah}\right)}{\sin^2\zeta}\frac{I_\nu}{\nu} \right]+
\frac{I_\nu}{\nu} \left \{3 n ^\alphah n^\betah \omega^{\zerohat}_{\alphah \betah} - n^\alphah e^\alpha_{\alphah;\alpha}\right. \nonumber\\
& & \left. -\frac{1}{\sin \zeta}\frac{\partial}{\partial \zeta}\left[ n^\alphah n^\betah
\left(\omega^{\threehat}_{\alphah \betah} - \omega^{\zerohat}_{\alphah \betah} n^{\threehat}\right)
  \right] - \frac{1}{\sin^2\zeta}\frac{\partial}{\partial \psi}\left[ n^\alphah n^\betah\left( n^{\twohat}\omega^{\onehat}_{\alphah \betah} - n^{\onehat}
\omega^{\twohat}_{\alphah \betah}\right)\right]\right \}\nonumber\\
& & =\frac{j_\nu - \alpha_\nu I_\nu}{\nu}.
\ee
This is roughly in flux conservative form except for the terms inside the braces, which we will now show evaluates to zero. We can rewrite the derivatives
\be
-\frac{1}{\sin \zeta}\frac{\partial \left(n^\alphah n^\betah\right)}{\partial \zeta} = \frac{n^{\threehat}(n^\alphah\delta^\betah_\zerohat+n^\betah\delta^\alphah_\zerohat)+n^\alphah\delta^\betah_\threehat +n^\betah\delta^\alphah_\threehat - 2 n^\threehat n^\alphah n^\betah}{\sin^2 \zeta}\nonumber\\
\frac{1}{\sin^2 \zeta}\frac{\partial (n^\alphah n^\betah)}{\partial \psi}=\frac{-n^\alphah n^\twohat \delta^\betah_\onehat
n^\alphah n^\onehat \delta^\betah_\twohat - n^\betah n^\twohat \delta^\alphah_\onehat + n^\betah n^\onehat \delta^\alphah_\twohat}{\sin^2 \zeta}.\label{eq:momderivs}
\ee
Defining
\be
\Theta & \equiv & 3 n ^\alphah n^\betah \omega^{\zerohat}_{\alphah \betah} - n^\alphah e^\alpha_{\alphah;\alpha}
 -\frac{1}{\sin \zeta}\frac{\partial}{\partial \zeta}\left[ n^\alphah n^\betah
\left(\omega^{\threehat}_{\alphah \betah} - \omega^{\zerohat}_{\alphah \betah} n^{\threehat}\right)
\right] \nonumber\\
& &  - \frac{1}{\sin^2\zeta}\frac{\partial}{\partial \psi}\left[ n^\alphah n^\betah\left( n^{\twohat}\omega^{\onehat}_{\alphah \betah} - n^{\onehat}
\omega^{\twohat}_{\alphah \betah}\right)\right],
\ee
we can evaluate the derivatives using equation (\ref{eq:momderivs}) while making use of equations (\ref{eq:riccisum}) and (\ref{eq:extraterm}) to find
\be
\Theta & = & \frac{n^\alphah}{\sin^2 \zeta}\left[
  n^\threehat (\omega^{\threehat}_{\alphah \zerohat} - n^\threehat \omega^{\zerohat}_{\alphah \zerohat})+
  (\omega^{\threehat}_{\alphah \threehat} - n^\threehat \omega^{\zerohat}_{\alphah \threehat})\right.\nonumber\\
  & & \left. +n_\onehat^2 \omega^{\twohat}_{\alphah \twohat} +n_\twohat^2 \omega^{\onehat}_{\alphah \onehat}-n^\onehat n^\twohat
  (\omega^{\onehat}_{\alphah \twohat} +\omega^{\twohat}_{\alphah \onehat})
  \right] +\frac{n^\betah}{\sin^2 \zeta}\left[
  n^\threehat (\omega^{\threehat}_{\zerohat \betah} - n^\threehat \omega^{\zerohat}_{\zerohat \betah})\right.\nonumber\\
  & & \left. +
  (\omega^{\threehat}_{\threehat \betah} - n^\threehat \omega^{\zerohat}_{\threehat \betah})
  + n_\onehat^2  \omega^{\twohat}_{\twohat \betah} + n_\twohat^2 \omega^{\onehat}_{\onehat \betah}-n^\onehat n^\twohat
  (\omega^{\onehat}_{\twohat \betah} + \omega^{\twohat}_{\onehat \betah}) \right]\nonumber\\
& & +\frac{n^\alphah n^\betah}{\sin^2 \zeta} (n^\onehat \omega^\onehat_{\alphah\betah} + n^\twohat \omega^\twohat_{\alphah\betah})
-n^\alphah e^\alpha_{\alphah;\alpha}.
  \ee
  We can simplify this significantly using the fact that Ricci matrices are antisymmetric in their first two indices.  Substituting $\gamma_{\zerohat \betah\gammah}=-\omega^\zerohat_{\betah\gammah}$ and $\gamma_{\ih \betah\gammah}=\omega^\ih_{\betah\gammah}$ and using antisymmetry in the first two indices, we find
\be
\Theta & = & \frac{n^\alphah}{\sin^2 \zeta}\left[
  n^\threehat (\gamma_{\threehat\alphah \zerohat} + n^\threehat \gamma_{\zerohat\alphah \zerohat})+
  (\gamma_{\threehat\alphah \threehat} + n^\threehat \gamma_{\zerohat\alphah \threehat})\right.\nonumber\\
  & & \left. +n_\onehat^2 \gamma_{\twohat\alphah \twohat} +n_\twohat^2 \gamma_{\onehat\alphah \onehat}-n^\onehat n^\twohat
  (\gamma_{\onehat\alphah \twohat} +\gamma_{\twohat\alphah \onehat})
  \right]\nonumber\\
& &+\frac{n^\alphah n^\betah}{\sin^2 \zeta} (n^\onehat \gamma_{\onehat\alphah\betah} + n^\twohat \gamma_{\twohat\alphah\betah})
-n^\alphah e^\alpha_{\alphah;\alpha}.
\ee
We also have
\be
-n^\alpha_{\alphah;\alpha} & = & -n^\alphah \omega^\betah_{\alphah\betah}\nonumber\\
 & = & n^\alphah\frac{\left[(\gamma_{\zerohat\alphah\zerohat}-
  \gamma_{\threehat\alphah\threehat}) (1-(n^\threehat)^2)-(\gamma_{\onehat\alphah\onehat}+
  \gamma_{\twohat\alphah\twohat})((n^\onehat)^2+(n^\twohat)^2)\right]}{\sin^2\zeta}\label{eq:extraterm},
\ee
which leaves us with
\be
\Theta & = &\frac{n^\alphah}{\sin^2 \zeta}\left[
  \gamma_{\zerohat\alphah \zerohat} + n^\threehat\gamma_{\zerohat \alphah \threehat}+n^\threehat\gamma_{\threehat\alphah \zerohat}
+ (n^\threehat)^2 \gamma_{\threehat \alphah \threehat} \right.\nonumber\\
& & \left. +n^\onehat \gamma_{\onehat\alphah \zerohat} +n^\onehat n^\threehat \gamma_{\onehat\alphah \threehat}
+n^\twohat \gamma_{\twohat\alphah \zerohat}+n^\twohat n^\threehat \gamma_{\twohat\alphah \threehat}\right]=0,
\ee
which is what we intended to show.  The last equality follows from the antisymmetry of the Ricci matrices in the first two indices. Hence, we arrive at equation (\ref{eq:fluxcon}).

\section{Phase Space Number Density Conservation}
\label{gammie}

\def\gdet{{\sqrt{-g}}}
\def\tf{{\tilde{f}}}
\def\dt{{\frac{\partial}{\partial t}}}
\def\dqi{{\frac{\partial}{\partial q^i}}}
\def\dpi{{\frac{\partial}{\partial p^i}}}
\def\dxi{{\frac{\partial}{\partial x^i}}}

The derivation in appendix \ref{davis} is formally correct but
somewhat obscures the physical principles.  Here we attempt to
elucidate how the equations follow from conservation of phase space
number density.  The streaming term (``Liouville operator'') in the
Boltzmann equation is derived by noting that (1) particle number is
conserved as particles flow through phase space, and (2) a Hamiltonian
flow is incompressible.  This leads to $df/d\lambda = 0$ in the absence of
interaction with matter.  Here we
will not apply (2) in order the write the equation in conservation
form.

Consider an arbitrary set of phase space coordinates $t, q^i, p^i$
(here, index up or down has no significance), and the fact that these are
labeled $p$ and $q$ do not imply that the $p,q$ are canonically conjugate.   

Define the (non-invariant) distribution function
\begin{equation}
\tf \equiv \frac{dN}{d^3q d^3p}
\end{equation}
(reserve $f \equiv dN/d^3x^i d^3p_i$ for the invariant distribution function).  Then one
can always write the streaming term in conservation form:
\begin{equation}\label{eq.boltzgen}
\dt \tf + \dqi (\dot{q}^i \tf) + \dpi (\dot{p}^i \tf) = 0.
\end{equation}
Here $\dot{}$ denotes a derivative with respect to $t$.
This equation is true always, independent of $p,q$, as long as they are not
pathological coordinate on the phase space \citep[see, e.g.][]{2008gady.book.....B}.

As a simple example, use the spirit of 
(\ref{eq.boltzgen}) to derive the particle conservation equation
in relativistic hydrodynamics.  Let $n_x \equiv dN/d^3x$.
Note that this is {\em not} the proper number density.  Then 
\begin{equation}
\del_t n_x + \del_i (n_x v^i)=0,
\end{equation}
where $v^i \equiv dx^i/dt = u^i/u^t$, where $u^\mu$ is the four-velocity.  
Rewrite this in terms of the proper density 
\begin{equation}
n = \frac{1}{u^t \gdet}\frac{dN}{d^3x} = \frac{n_x}{\gdet u^t} 
\end{equation}
This is invariant because $u^t \gdet d^3x$ is invariant.  Then
\begin{equation}
\del_t (\gdet u^t n) + \del_i (\gdet u^i n)=0,
\end{equation}
which is the usual covariant continuity equation, with all the 
$u^\alpha$'s in the right places.

Now specialize to phase space coordinates $x^i, \zeta, \psi, \nu$.   The
Boltzmann becomes
\begin{equation}\label{eq.boltzspec}
\dt  \tf + \frac{\partial}{\partial x^i} (\dot{x}^i \tf) + 
\frac{\partial}{\partial \zeta} (\dot{\zeta} \tf) +
\frac{\partial}{\partial \psi} (\dot{\psi} \tf) +
\frac{\partial}{\partial \nu} (\dot{\nu} \tf)  = 0.
\end{equation}
To make this complete we need only (1) deduce the relation between $\tf$
and $I_\nu$ defined in the tetrad frame, and (2) evaluate the derivatives
such as $\dot{\zeta}$ using $d/dt = (1/k^t) d/d\lambda$.

Recall that 
\begin{equation}
\tf = \frac{dN}{d^3x d\zeta d\psi d\nu}
\end{equation}
and
\begin{equation}
I_\nu = \frac{dE}{dA' dt' d\Omega d\nu} = h \nu \frac{dN}{dA' dt' d\Omega d\nu}
\end{equation}
where the $'$ refers to values in the tetrad frame.  Then using the
invariance of
\begin{equation}
k^{(t)} d^3x' = \nu c dt' dA' = \sqrt{-g} k^t d^3x 
\end{equation}
(notice that this is $k^t$ in the coordinate frame) and the definition
\begin{equation}
\sin\zeta d\zeta d\psi = d\Omega 
\end{equation}
it is easy to show that
\begin{equation}
\tf = \frac{k^t}{\nu} \, \gdet \, \sin\zeta \, \frac{I_\nu}{h\nu}.
\end{equation}
This explains why the combination $I_\nu/\nu$ appears in (\ref{eq:fluxcon}).

We can immediately assemble the first four terms in (\ref{eq.boltzspec})
into:
\begin{equation}
\frac{\del}{\del x^\mu} \left(
\gdet \sin\zeta n^\mu 
\frac{I_\nu}{h\nu} 
\right).
\end{equation}
Here $n^\alpha = k^\alpha/\nu$.
Notice that we can divide this term by $\sin\zeta \gdet/h$ to get agreement with
expressions in the text, since $(v^\mu)_{;\mu} = \del_\mu (\gdet v^\mu)/\gdet$.

Now we just need to evaluate the momentum terms, which involve
time derivatives of the momentum space coordinates, which we already worked out in appendix
\ref{davis}. For example, $\cos\zeta = n^{(3)} = k^{(3)}/k^{(0)}$.

Since equation (\ref{eq:connect}) tells us how to take the derivatives of each
component of $k$ in the tetrad basis, we are basically done:
\begin{equation}
\dot{k}^{(3)} = - k^{(a)} k^{(b)} \omega^{(3)}_{(a)(b)}
\end{equation}
(now $\dot{}$ denotes a derivative with respect to $\lambda$)
\begin{equation}
\dot{k}^{(0)} = - k^{(a)} k^{(b)} \omega^{(0)}_{(a)(b)}
\end{equation}
so
\begin{equation}
\sin\zeta \, \dot{\zeta} = 
\frac{k^{(a)} k^{(b)} }{k^{(0)}} \omega^{(3)}_{(a)(b)}
-\frac{k^{(3)} k^{(a)} k^{(b)}}{k^{(0) 2}} \omega^{(0)}_{(a)(b)}
\end{equation}
or
\begin{equation}
\dot{\zeta} = \nu \frac{n^{(a)} n^{(b)} }{\sin\zeta} 
\left(\omega^{(3)}_{(a)(b)} - \omega^{(0)}_{(a)(b)} n^{(3)} \right).
\end{equation}
Putting everything together, and not forgetting the $k^t$ in the
denominator to convert from $d/dt$ to $d/d\lambda$, the $\zeta$ term 
in (\ref{eq.boltzspec}) is
\begin{equation}
\frac{\partial}{\partial \zeta} (\dot{\zeta} \tf)=
\frac{\del}{\del\zeta} \left(
n^{(a)} n^{(b)} 
\left(\omega^{(3)}_{(a)(b)} - \omega^{(0)}_{(a)(b)} n^{(3)} \right)
\frac{I_\nu}{h\nu} \gdet
\right).
\end{equation}
which agrees with (\ref{eq:fluxcon}), apart from a factor of $\sin\zeta \gdet/h$.

The $\nu$ term requires that we evaluate
\begin{equation}
\dot{\nu} = \dot{k^{(0)}} = 
- k^{(a)} k^{(b)} \omega^{(0)}_{(a)(b)}
\end{equation}
so we can immediately write 
\begin{equation}
\frac{\partial}{\partial \nu} (\dot{\nu} \tf) =- \frac{\del}{\del\nu} \left(
n^{(a)} n^{(b)} \omega^{(0)}_{(a)(b)}
\gdet\,\sin\zeta \, \frac{I_\nu}{h}.
\right).
\end{equation}
This agrees with (\ref{eq:fluxcon}) apart from a factor of $\sin\zeta \gdet/h$.

Finally, the $\psi$ term can be evaluated using
\begin{equation}
\tan \psi = \frac{k^{(2)}}{k^{(1)}}.
\end{equation}
Then 
\begin{equation}
\sec^2\psi \, \dot{\psi} = 
- \frac{ k^{(a)} k^{(b)}}{k^{(1)}} \omega^{(2)}_{(a)(b)}
+ \frac{ k^{(2)} k^{(a)} k^{(b)}}{k^{(1) 2}} \omega^{(1)}_{(a)(b)}
\end{equation}
which can be rewritten
\begin{equation}
\dot{\psi} = 
\nu \frac{n^{(a)} n^{(b)}}{\sin\zeta} \left(
\sin\psi \omega^{(1)}_{(a)(b)}
- \cos\psi \omega^{(2)}_{(a)(b)} \right)
\end{equation}
Putting everything together yields
\begin{equation}
\frac{\del}{\del\psi} \left(
n^{(a)} n^{(b)} \left(
\sin\psi \omega^{(1)}_{(a)(b)}
- \cos\psi \omega^{(2)}_{(a)(b)} \right)
\gdet \, \frac{I_\nu}{h\nu}
\right).
\end{equation}
which looks slightly different from (\ref{eq:fluxcon}), but can be rewritten using the
definition of $\zeta,\psi$ as
\begin{equation}
\frac{\partial}{\partial \psi} (\dot{\psi} \tf)=\frac{\del}{\del\psi} \left(
\frac{ n^{(a)} n^{(b)} }{\sin\zeta} \left(
n^{(2)} \omega^{(1)}_{(a)(b)}
- n^{(1)} \omega^{(2)}_{(a)(b)} \right)
\gdet \, \frac{I_\nu}{h \nu}
\right).
\end{equation}
This again agrees up to a factor of $\sin\zeta \gdet/h$ and we verify that equation~(\ref{eq.boltzspec}) is equivalent to equation~(\ref{eq:fluxcon}).

\section{Second Moment Relation}
\label{secondmoment}

Equation (\ref{eq:momentsimp}) follows from multiplying equation (\ref{eq:fluxconalt}) by $n_\beta$ to obtain
\be
(n^\alpha n_\beta I_\nu)_{;\alpha} & + &
\frac{\partial}{\partial \nu}\left(\fnu n_\beta I_\nu \right)
-\frac{1}{\sin \zeta}\frac{\partial}{\partial \zeta}\left( \fzeta n_\beta I_\nu\right)
+\frac{\partial}{\partial \psi}\left(\fpsi n_\beta I_\nu \right)\nonumber\\
& - & \frac{n_\beta \fnu}{\nu}+\frac{\fzeta}{\sin \zeta}\frac{\partial n_\beta }{\partial \zeta}
-\fpsi \frac{\partial n_\beta }{\partial \psi}-n^\alpha n_{\beta;\alpha}
=n_\beta \left(j_\nu - \alpha_\nu I_\nu\right),
\ee
and demonstrating that
\be
-\frac{n_\beta \fnu}{\nu}+\frac{\fzeta}{\sin \zeta}\frac{\partial n_\beta }{\partial \zeta}
-\fpsi \frac{\partial n_\beta }{\partial \psi}-n^\alpha n_{\beta;\alpha}=0.
\ee
Since the metric is independent of momentum coordinates and its covariant derivative vanishes, this is equivalent to showing that
\be
\Xi^\beta \equiv -\frac{n^\beta \fnu}{\nu}+\frac{\fzeta}{\sin \zeta}\frac{\partial n^\beta }{\partial \zeta}
-\fpsi \frac{\partial n^\beta }{\partial \psi}-n^\alpha n^\beta_{\;;\alpha}=0.
\ee
Using equation (\ref{eq:fluxes}) along with
\be
n^\alpha n^\beta_{\;;\alpha} & = & n^\alpha n^\betah e^\beta_{\betah;\alpha}\nonumber\\
& = & n^\alphah n^\betah e^\beta_\gammah \omega^\gammah_{\alphah \betah}\nonumber\\
& = &  \frac{n^\alphah n^\betah}{\sin^2 \zeta} \left[\left(\omega^\zerohat_{\alphah \betah} e^\beta_\zerohat +
  \omega^\threehat_{\alphah \betah} e^\beta_\threehat\right)(1-n_\threehat^2)
  +\left(\omega^\onehat_{\alphah \betah} e^\beta_\onehat +\omega^\twohat_{\alphah \betah}
  e^\beta_\twohat\right)(n_\onehat^2+n_\twohat^2) \right]
\ee
and evaluating the derivatives with respect to $\psi$ and $\zeta$,
\be
-\frac{1}{\sin \zeta}\frac{\partial n^\beta}{\partial \zeta} & = & \frac{e^\beta_\zerohat n^\threehat +e^\beta_\threehat +n^\threehat n^\beta}{\sin^2 \zeta}\nonumber\\
\frac{\partial n^\beta}{\partial \psi} & = & -n^\twohat e^\beta_\onehat +n^\onehat e^\beta_\twohat
\ee
we obtain
\be
\Xi^\beta & = & n^\alphah n^\betah \left[\omega^\threehat_{\alphah\betah}\left(e^\beta_\zerohat
  -n^\threehat n^\beta+e^\beta_\threehat n_\threehat^2\right)\right.\nonumber\\
  & &\left. +\omega^\zerohat_{\alphah\betah}\left(-e^\beta_\threehat +n^\beta
  -e^\beta_\zerohat\right)+\omega^\onehat_{\alphah\betah}\left(n^\onehat e^\beta_\zerohat +n^\onehat n^\threehat e^\beta_\threehat-n^\onehat n^\beta
  \right)\right.\nonumber\\
& & \left.\omega^\twohat_{\alphah\betah}\left(n^\twohat e^\beta_\zerohat +n^\twohat n^\threehat e^\beta_\threehat-n^\twohat n^\beta
  \right)  \right].
\ee
Making liberal use of equation (\ref{eq:riccisum}) one can easily show that $\Xi^\beta=0$.

\clearpage

\newpage
\bibliographystyle{apj}
\bibliography{local}

\end{document}